\documentstyle[psfig,preprint,aps]{revtex}

\def\sss{\scriptscriptstyle}
\begin{document}
\draft

\preprint{Manuscript \# LD6759}

\title{\vspace{.5cm} Vortex-nucleus interaction and pinning forces in neutron
stars} 
 
\author{P. M. Pizzochero$^{a,b}$, L. Viverit$^{a}$ and R. A. Broglia$^{a,b,c}$} 

\address{
     \vspace{.3cm} 
     {\setlength{\baselineskip}{18pt}
     $^a$ Dipartimento di Fisica, Universit\`a degli Studi di Milano,\\
     Via Celoria 16, 20133 Milano, Italy.\\
     $^b$ INFN, Sezione di Milano, Via Celoria 16, 20133 Milano, Italy.\\
     $^c$ The Niels Bohr Institute, University of Copenhagen,\\
     Blegdamsvej 17, 2100 Copenhagen \O, Denmark.\\} }

\date{Revised version}

\maketitle

\begin{abstract}
\noindent 
We calculate the  force that pins vortices in the neutron superfluid to  nuclei
in the inner crust of rotating neutron stars, relying on a detailed microscopic
description of both the vortex radial profile and the inner crust nuclear
structure. The contribution to the pinning energy from pair condensation  is
estimated in the local density approximation with realistic nucleon-nucleon
interactions. The kinetic contribution, not consistently included in previous
approaches, is evaluated in the same approximation and found to be relevant.
The vortex-nucleus interaction turns out to be attractive for stellar 
densities greater than $\sim 10^{13}$ g/cm$^{3}$. In this region, we find
values for the pinning force which are almost one order of magnitude lower than
the ones obtained so far. This has  direct consequences on the critical
velocity differences for vortex depinning. \\ 
\end{abstract}

\pacs{PACS number(s): 26.60.+c}

\newpage

The calculation of the interaction energy between a vortex and a nucleus has
been of  high concern since the vortex pinning model  was proposed by Anderson
and Itoh \cite{And} to explain pulsar glitches, that is sudden spin-ups in the
neutron star rotation.  The idea is to calculate the difference in energy
between  a configuration with the nucleus outside the vortex core and one with 
the nucleus at the center of  the core. This is done by taking the
configuration of a vortex alone as the one of zero energy, and then calculating
the energies of the two configurations with the nucleus present. The first
estimates \cite{Alp} considered only the difference in pairing
condensation energy, calculated in a crude model with uniform densities for
both nuclear and vortex matter. As made clear by Epstein and Baym  \cite{EB},
however, the difference in energy between a vortex alone and one with a nucleus
comes from two contributions, one of which is kinetic and the other
condensational.  These authors also introduced a realistic density profile for 
the nuclei present in the neutron star crust, which has a relevant effect on
the results for the pairing energies. To date, their treatment is the most
refined available in the literature, although, as discussed later, they only
use the condensational contribution to evaluate the pinning energy. 

The point of view of Epstein and Baym \cite{EB} was to use the Ginzburg-Landau
approximation to evaluate the pairing properties of the superfluid crust. In
this scenario, the core radius was taken to be $\xi_{\sss GL}$, the
Ginzburg-Landau order parameter.  The conditions of applicability of the
Ginzburg-Landau theory, however, are far from satisfied in the case under
discussion.  Indeed, the neutron star crust is practically a zero-temperature
case ($T\sim 0.01$ MeV), while for the Ginzburg-Landau approach to be valid,
the temperature of the system should be close to the transition one ($T_{c}
\sim 0.5 $ MeV). Moreover, the density variations due to the presence of the
nucleus are quite steep, which is also in contrast with the requirements of the
Ginzburg-Landau  theory. As a matter of fact,  Epstein and Baym must rescale
their results for the  Ginzburg-Landau coherence lengths by factors in the
range $ 2 - 12 $, in order to reproduce experimental condensation energies for
ordinary nuclei. 

For  these reasons we felt the need to change the theoretical framework and use
a  more realistic approach to treat the radial dependence of the pairing gap in
the inner crust of neutron stars. Our model is based on the local density
approximation to evaluate the pairing properties of the system \cite{BS}. This
approach, when applied to ordinary finite nuclei, gives realistic values for
their condensation energies \cite{Kuc}.  Application of this model to the case
of the inner crust of neutron stars, where a lattice of neutron rich nuclei
(described in terms of Wigner-Seitz cells) is permeated by a gas of unbound
superfluid neutrons, can be found in Ref.~\cite{Bro}.  A full BCS treatment of
the problem, although more satisfactory, would bring about many difficulties,
due to the different symmetries and yet comparable dimensions  of the nucleus
and the vortex core. 

Superfluid matter in a straight vortex moves with a velocity field
\begin{equation}
{\bf v}({\bf x})=\frac{\hbar}{2m_{\sss N}r} {\bf e}_{\vartheta} \ ,  \label{vel}
\end{equation}
where r is the radial distance of the point {\bf x} to the vortex axis, 
$m_{\sss N}$ 
is the nucleon mass and ${\bf e}_{\vartheta}$ is the tangent unit vector.
>From this equation we can readily see the need for a layer of normal matter,
called vortex core, surrounding the axis and co-rotating with the solid crust.
This is so since the curl of the velocity field of a superfluid has to be zero
everywhere. The field we are considering satisfies this condition at every
point but on the axis. This singular behaviour can be avoided by assuming that,
along the axis, matter is not superfluid. This point can be understood also in
another way. Eq.~(\ref{vel}) states  that the velocity and the kinetic
energy 
density of the superfluid tends to infinity as the axis is approached. This is
clearly impossible, thus indicating that at some point close to the axis
neutron matter has to undergo a transition to a normal state, where it can be
assumed to be static in a frame where the nuclear lattice is at rest. 

This is the point of view we took to define the radius core. In this case, the
distance where the transition occurs can be obtained  equating the kinetic
energy density, due to the rotation around the axis, to the condensation energy
per unit volume. Closer to the axis the kinetic term  increases rapidly, making
it energetically unfavorable for matter to remain superfluid. The kinetic
energy per unit volume is 
\begin{equation}
{\cal E}_{kin}=\frac{\hbar^{2} n}{8m_{\sss N}r^{2}} \ ,       \label{ecin}
\end{equation}
where $n= n(r)$ is the superfluid particle (neutron) density. Due to the
superfluid state, a unit volume of matter has an energy lower by 
\begin{equation}
{\cal E}_{cond}=-\frac{3\Delta^{2} n}{8\varepsilon_{\sss F}} \ ,  \label{econd}
\end{equation}
compared to a unit volume of normal matter. Here $\Delta = \Delta(r)$ is the
energy gap calculated in the semiclassical approximation, and $\varepsilon_{\sss
F} = \varepsilon_{\sss F}(r)$ is the local Fermi energy. In the   local density
approximation, the different quantities depend parametrically on $r$ via the
local Fermi momentum (see Ref.~\cite{Bro}).  Equating
Eqs.~(\ref{ecin}) and (\ref{econd}) to zero, one gets an equation in $r$, whose solution
is the transition radius, $R_{t}$. This argument  can be readily generalized to
the case in which a nucleus is set at the center of the vortex core, thus
modifying its structure. Numerical calculations, using a realistic neutron
density profile as given by Negele and Vautherin \cite{NV}, were performed to
obtain the shape of the core. In this case, the transition radius $R_{t} =
R_{t}(z)$ will depend also on the coordinate $z$, due to the spherical symmetry
of the nucleus.   In the actual calculations, we took also into account  the
density variation of the neutron superfluid due to the centrifugal potential
induced by the rotation, as follows  from the local density
approximation.

As said before, the vortex alone was considered as the zero energy state.
Setting a nucleus within the flow changes the density profile and the velocity
field, thus causing a variation in the kinetic and condensation energy.
Depending on where the nucleus is placed, the energies will be  modified by a
different amount. We considered the two cases of a nucleus right at the center
of the core (case~I) and just barely out of it (case~II). 

In case~I, the kinetic term was obtained by a numerical integration of the
kinetic energy density, given by Eq.~(\ref{ecin}).  The condensation energy was
obtained via an numerical integration of Eq.~(\ref{econd}) over the volume occupied
by the superfluid. In both cases, the realistic density profiles were used. We
point out that Epstein and Baym \cite{EB} neglect the kinetic contribution in this
case, while our results shows that it is relevant. Incidentally, a simple
calculation based on their approach (and in the simplified scenario of purely
axial symmetry, i.e. with a ``cylindrical'' nucleus) gives a kinetic effect of
magnitude comparable to ours. 

Epstein and Baym \cite{EB} gave a good estimate of the kinetic energy when the
nucleus is out of the vortex core, and we took that as the appropriate value.
To find the condensation energy term in case~II, we proceeded  as before by
numerical integration. We point out that Epstein and Baym, after calculating
the kinetic contribution, do not include it in the evaluation of the pinning
energies. In this sense, their results  effectively include only the pairing
contribution, calculated in the Ginzburg-Landau approximation. 

Subtraction of the energy of a nucleus outside the core and that of one inside,
yields the pinning energy, $E_{pin}$. The pinning force, $F_{pin}$, is defined
as $E_{pin}$ divided by the minimum distance between the nucleus and the vortex
axis. This was taken to be $R_{t}+R_{\sss N}$, where $R_{\sss N}$ is the
nuclear radius. 

We performed our calculations for different zones in the inner crust of the
neutron star. The physical properties of these zones were obtained by Negele
and Vautherin \cite{NV} and we report them in Table~1. The calculations where
done using different nucleon-nucleon residual interactions, namely Argonne's
potential \cite {Bal} and Gogny's effective interaction \cite{Bro}, and with
the nucleon effective mass varying with density . Incidentally, it turns out
that setting the effective mass  equal to that of a free nucleon does not
change the results significantly. 

In the Table~2 and Table~3  we report  the results obtained. The transition
radius $R_{t}$ is the core radius of the vortex alone. In order to compare
kinetic and pairing contributions, we give the values for $\Delta E_{kin} = 
E_{kin,out} - E_{kin,in}$ and $\Delta E_{cond} = E_{cond,out} - E_{cond,in}$,
so that the pinning energy is $E_{pin} = \Delta E_{kin} + \Delta E_{cond}$. The
subscript `in' refers to the state in which the nucleus is at the center of a
vortex core, and `out' to the case of a nucleus whose center is at a distance
$R_{t}+R_{\sss N}$ from the core axis. When the pinning energy is positive, the
vortex pins to nuclei. When the pinning energy is negative, the vortex tends to
avoid nuclei in its path through the lattice. We refer to this scenario as
threading (or interstitial pinning). Only in the pinning case,  can we calculate
the pinning force as just described. In the threading case, instead, it is much
easier for the vortex to move through the nuclear array, and the pinning force
is orders of magnitude smaller than the values one would obtain from $E_{pin}$
(cf. Ref.~\cite{LE}).  
 
A general look at the results shows that there is pinning on nuclei for
densities greater than $\sim 10^{13}$ g/cm$^{3}$. This general trend is in
agreement with what has been  so far obtained in the literature. As already
mentioned, the kinetic energy contributions are relevant, as can be seen from
the relative values of $\Delta E_{kin}$ and $\Delta E_{cond}$. In particular,
due to the interplay between the spherical geometry of the nucleus and the
cylindrical geometry of the vortex, the kinetic energy difference can be also
negative. The Argonne and Gogny cases are quite similar, although Gogny gives
pinning only at slightly larger densities. The fact that these very different
interactions (Argonne is a bare nuclear potential, Gogny is an effective 
interaction) give results for the pinning that agree within a factor of two is
gratifying, since the choice of the nucleon-nucleon interaction to be used
in the calculations discussed here is an open and controversial issue. 

We now compare our results with those obtained by other authors. In Table~4 we
report the values for the pinning energies obtained by Epstein and Baym
\cite{EB}, as well as the results for the pinning force calculated by Link and
Epstein \cite{LE} from those energies. We remind that the pinning energies of
Epstein and Baym are only condensational (i.e., they correspond to the term
$\Delta E_{cond}$). We notice how their pairing energy differences are much
larger than ours. This is due to the fact that, in order to reproduce
experimental condensation energies for ordinary nuclei in the   Ginzburg-Landau
approach, they must divide their coherence lengths by factors in the range $ 2
- 12 $ (depending on the pairing gaps they use).  In turn, this amounts to
multiplying the condensation energies by factors in the range $ 4 - 144$.
Finally, after averaging the results obtained from  two sets of pairing gaps 
(`Takatsuka' and `Chen {\it et al.}' gaps \cite{EB}),
they obtain the `best-estimates' for the pairing energy difference reported in
Table~4. Numerically, however, the kinetic contribution included by us
partially makes up for the difference, since it presents  relevant positive
values at larger densities. 

To complete the comparison between our results and those obtained by 
Epstein and Baym in the Ginzburg-Landau approximation, we first observe that 
the pairing gaps calculated in neutron matter with the Argonne interaction 
\cite{Bal} and those calculated by Takatsuka \cite{T}  are practically the same 
in the density range corresponding to the inner crust. Therefore, it is 
instructive to compare the difference in pairing energy $ \Delta E_{cond} $ 
obtained in the present paper with the Argonne potential, and that obtained by 
Epstein and Baym with the Takatsuka gaps (which can be deduced from table~4 of 
Ref.~\cite{EB}). These results are reported in Table~5. The two sets of values
differ by one order of magnitude, thus confirming the striking difference
between the two approaches. We have already discussed how the local density 
approximation is expected to be a better approach than the Ginzburg-Landau one
for the situation under study.

>From a general look to the previous results, we see that our treatment gives
pinning forces that are  smaller than those obtained in the previous approaches
by almost one order of magnitude. We point out that having too large values for
the pinning force has been one of the problems of the vortex pinning model. In
this sense, the results of our approach seem to go in the right direction.

In conclusion, we have proposed a  microscopic model to calculate the 
vortex-nucleus interaction in the inner crust of rotating neutron stars. We 
have treated the pairing energies in a semiclassical approximation, which is
 better suited to deal with the system under discussion
than the Ginzburg-Landau approach followed so far. We have also
included the kinetic contribution to the pinning energy, which turns out to be
relevant. We have used realistic density profiles for the Wigner-Seitz cells
and different realistic nucleon-nucleon interactions to test their influence.
We have defined the radius of the vortex core and the density profile of the
rotating superfluid in a way which is consistent with the semiclassical
approach followed. In particular, we have not introduced any arbitrary scaling
factor in our model. We have obtained results that differ
by almost one order of magnitude from those obtained in previous less
refined approaches. These results are likely to have important effects in
relation to pulsar glitches. For example,  critical velocity differences for
depinning are directly related to the pinning forces. These applications,
however, are beyond the scope of the present work. \\

\newpage

\noindent
TABLE 1 -- Physical parameters of the four regions in the inner crust. The
values are taken from Negele and Vautherin \cite{NV}. The baryon densities,
$\rho_b$, of the four zones are given in g/cm$^{3}$, the densities of the free
neutron gas, $n_{n{\sss G}}$, in fm$^{-3}$ and the radii of the nuclei, $R_{\sss
N}$, and those of the Wigner-Seitz cells, $R_{\sss WS}$, in fm. \\

\begin{center}
\begin{math}
\begin{array}{|c||r|r|r|r|}
\hline
\ {\sl Zone} \  &  1 & 2 & 3 & 4  \\
\hline \hline
\ \rho_b & \ 1.51\times 10^{12} & \ 9.55\times 10^{12} & \ 3.39\times 10^{13} &
\ 7.76\times 10^{13} \\ 
\ n_{n{\sss G}} & \ 4.79\times 10^{-4} & \ 4.68\times 10^{-3} & \ 1.82\times 10^{-2} &
\ 4.37\times 10^{-2} \\ 
\ R_{\sss N} & 6.0 & 6.73 & 7.32 & 6.72 \\
\ R_{\sss WS} & 44.0 & 35.5 & 27.0 & 19.4 \\
\hline
\end{array}
\end{math}
\end{center}

\vskip 2cm

\noindent
TABLE 2 -- Results of the calculation with the Argonne interaction. The radii
of the vortex core, $R_t$, are given in fm, the energies  in MeV, while the
pinning forces are in MeV/fm. As explained in the text, the pinning forces are
given only for positive pinning energies, since in the threading regime they do
not derive from the values of $E_{pin}$ \cite{LE}. \\

\begin{center}
\begin{math}
\begin{array}{|c||r|r|r|r|}
\hline
\ {\sl Zone} \ & 1 & 2 & 3 & 4  \\
\hline \hline 
\ R_{t} & 3.87 & 2.93 & \ \ 3.62 & \ \ 7.02 \\
\  \Delta E_{kin}  & \ -2.59 &  0.52 & 5.36 & 1.25 \\ 
\ \Delta E_{cond}  & -0.31 & \ -0.42 & 0.63  & 2.69 \\ 
\ E_{pin} & -2.90 & 0.10 & 5.99 & 3.94 \\
\  F_{pin}  &  & 0.01 & 0.55 & 0.29 \\
\hline
\end{array}
\end{math}
\end{center}

\newpage

\noindent
TABLE 3 -- Results of the calculation with the Gogny interaction. The radii of
the vortex core, $R_t$, are given in fm, the energies are in MeV and the pinning
forces  in MeV/fm.

\begin{center}
\begin{math}
\begin{array}{|c||r|r|r|r|}
\hline
\ {\sl Zone} \ & 1 & 2 & 3 & 4  \\
\hline \hline 
\ R_{t} & 2.88 & 2.44 & 2.82 & 5.13 \\
\  \Delta E_{kin}  & \ -3.90  & \ -0.36 & 6.32 & 5.89 \\ 
\ \Delta E_{cond}  & -0.28 & -1.24 & \ -0.37 & \ \ 1.60 \\ 
\ E_{pin} & -4.18 & -1.60 & 5.95 & 7.49 \\
\ F_{pin} &  &  & 0.59 & 0.63 \\ 
\hline
\end{array}
\end{math}
\end{center}

\vskip 1.5cm

\noindent
TABLE 4 -- Results from the Ginzburg-Landau approximation. The pinning energies
are taken from Epstein and Baym  \cite{EB}, the pinning forces from Link and
Epstein \cite{LE}. The  energies are given  in MeV and the forces   in MeV/fm.

\begin{center}
\begin{math}
\begin{array}{|c||r|r|r|r|}
\hline \ {\sl Zone} \ & 1 & 2 & 3 & 4 \\
\hline \hline
\  E_{pin} & \ \ -4.4  & 0.4  & \ 15.0  & \ \ 9.0  \\
\ F_{pin} &  & \  0.11 & 3.6 & 1.9 \\
\hline
\end{array}
\end{math}
\end{center}

\vskip 1.5cm

\noindent
TABLE 5 -- Difference in pairing energy $ \Delta E_{cond} $, obtained in this
paper in the local density approximation with the Argonne interaction, and
obtained by Epstein and Baym \cite {EB} in the Ginzburg-Landau approximation
with the Takatsuka gaps \cite{T}. The  energies are given  in MeV.

\begin{center}
\begin{math}
\begin{array}{|c||r|r|r|r|}
\hline \ {\sl Zone} \ & 1 & 2 & 3 & 4 \\
\hline \hline
\  {\rm Argonne \ } & \ -0.31  & \ -0.42  & \ 0.63  & \ \ 2.69  \\
\ {\rm Takatsuka} \ & -4.2  & \  -4.8 & \ 11.9 & 17.1 \\
\hline
\end{array}
\end{math}
\end{center}

\end{document}